\documentclass[superscriptaddress,
 amsmath,amssymb,
 aps,
 prl,
 reprint,
 floatfix
]{revtex4-2}

\usepackage{graphicx}
\usepackage{dcolumn}
\usepackage{array}
\usepackage{physics}
\usepackage{amsfonts, amsmath}
\usepackage{dsfont}
\usepackage{amsthm}
\usepackage{bm}
\usepackage{hyperref}
\usepackage{bbold}
\usepackage{color}
\usepackage[normalem]{ulem}
\usepackage{tabularx}
\usepackage{tikz}
\usetikzlibrary{arrows.meta,calc}

\newcommand{\proj}[1]{\lvert #1\rangle\langle #1\rvert}

\begin{document}
\definecolor{navy}{RGB}{46,72,102}
\definecolor{pink}{RGB}{219,48,122}
\definecolor{grey}{RGB}{184,184,184}
\definecolor{yellow}{RGB}{255,192,0}
\definecolor{grey1}{RGB}{217,217,217}
\definecolor{grey2}{RGB}{166,166,166}
\definecolor{grey3}{RGB}{89,89,89}
\definecolor{red}{RGB}{255,0,0}

\title{Ultimate Information Rate for Quantum Sensing under Multilevel Relaxation}

\author{Changhun Oh}
\email{changhun0218@gmail.com}
\affiliation{Department of Physics, Korea Advanced Institute of Science and Technology, Daejeon 34141, Korea}
\author{Seok Hyung Lie}
\email{seokhyung@unist.ac.kr}
\affiliation{Department of Physics, Ulsan National Institute of Science and Technology, Ulsan 44919, Korea}
\author{Youngrong Lim}
\email{sshaep@gmail.com}
\affiliation{Department of Physics, Chungbuk National University, Cheongju 28644, Korea}

\begin{abstract}
We determine the ultimate information rate of a relaxing multilevel quantum sensor under unrestricted adaptive control and construct an explicit strategy that attains it. 
We consider sensing a weak field that couples the ground state to several excited states that decay back to the ground state, a setting that reduces to amplitude-damping sensing for a single excited state.
We show that the ultimate information rate is exactly the mean population lifetime of the bright state coupled to the ground state by the signal. 
Although the sensing dynamics generally involve several decay modes, an explicit rank-one protocol that monitors the ground--bright coherence using fresh meters and classical feedback attains this rate while keeping its finite-time information deficit bounded by a constant.
For independently relaxing identical sensors, independent local monitoring attains the sum of their individual optimal rates, so intersensor entanglement and joint quantum error correction cannot increase the asymptotic rate. 
Consequently, the free population-decay curve provides an operational measure of the optimal sensing performance under arbitrary adaptive control.
\end{abstract}

\maketitle

Quantum metrology uses coherent dynamics, entanglement, and control to enhance the response of a sensor to weak signals \cite{BraunsteinCaves1994,Giovannetti2004,Giovannetti2006,Degen2017}. In realistic devices, the same coherence that carries the signal is continuously degraded by relaxation and decoherence, limiting the information that can be accumulated \cite{Huelga1997,Chaves2013,Kolodynski2013,DemkowiczMaccone2014,Alipour2014,Smirne2016}. The resulting challenge is to determine the maximum information generated per unit time, and how it scales with the number of sensors, when coherent control, noiseless quantum memories, intermediate measurements, feedback, quantum error correction~(QEC), and intersensor entanglement are all available \cite{Dur2014,Arrad2014,Kessler2014,HerreraMarti2015,Lu2015,Sekatski2017,Zhou2018,Layden2019,ZhouJiang2020,Kurdzialek2023,Demkowicz2017,WanLasenby2022}.

Intuitively, relaxation creates a competition between signal generation and information loss. 
A longer interrogation produces a larger sensor response, but also gives the environment more time to erase it. 
Continuous monitoring can extract information into a classical record while retaining sensitivity in the conditional sensor state~\cite{GammelmarkMolmer2013,Gammelmark2014,KiilerichMolmer2014,Catana2015,KiilerichMolmer2016,Albarelli2017,Albarelli2018,Ma2019,Yang2023,Yang2026}. Measurement-based feedback can then control the subsequent dynamics~\cite{WisemanMilburn1993}.
Experimentally, related techniques have enabled sequential weak readout of single-nuclear-spin dynamics~\cite{Cujia2019,Pfender2019,Meinel2022} and enhanced sensing using a quantum memory~\cite{Zaiser2016,Arunkumar2023}.
For an amplitude-damping qubit, the ultimate adaptive limit and an explicit strategy attaining it are now understood~\cite{Demkowicz2017,Kurdzialek2025}.

Beyond a single two-level transition, coherent access to multiple internal levels allows a sensor to exploit transitions and superpositions with different signal couplings and lifetimes~\cite{Mamin2014,Chalopin2018,Evrard2019,Ma2026}.
Under relaxation, weakly coupled but long-lived components can contribute substantially to the information accumulated over time, even when their initial response is small.
The central question is how these contributions jointly determine the ultimate information rate and what control resources are needed to attain it.

In this work, we determine the ultimate information rate under unrestricted adaptive control for a multilevel sensor whose excited states undergo Markovian relaxation to a common ground state.
A weak field couples the ground state to a bright superposition of excited states, while relaxation can act on different superpositions with unequal decay rates~\cite{Agarwal2000,Bennett2021,Evangelou2026}.
We show that the entire relaxation geometry enters this rate through a simple quantity, the mean population lifetime of the bright state.
The free population-decay curve thus quantifies the combined sensing contribution of all signal-coupled decay modes.
We further identify the rate-optimal response and show that it can be maintained by a fixed weak measurement of the ground--bright coherence and classical feedback, without separately addressing the decay modes or retaining a quantum memory between readouts.
This protocol attains the asymptotic rate allowed by the general adaptive bound for Markovian metrology~\cite{FujiwaraImai2008,Escher2011,Demkowicz2012,Demkowicz2017,WanLasenby2022}.

This single-sensor result also determines the ultimate rate for multiple independently relaxing sensors. 
For $n$ identical sensors with independent relaxation, the maximum quantum Fisher information (QFI) $F_{\rm opt}^{(n)}(T)$ attainable in time $T$ by an unrestricted joint adaptive strategy satisfies
\begin{align}
\lim_{T\to\infty}\frac{F_{\rm opt}^{(n)}(T)}{T}&=nR,\qquad R:=\bra bG^{-1}\ket b.
\label{eq:mainrate}
\end{align}
Here, $\ket b$ is the bright state and $G$ is the relaxation operator. 
Applying the monitoring protocol independently to each sensor attains this unrestricted limit. 
Intersensor entanglement and joint QEC, therefore, cannot improve the asymptotic information rate.

\medskip

\begin{figure}[t]
\centering
\includegraphics[width=0.95\columnwidth]{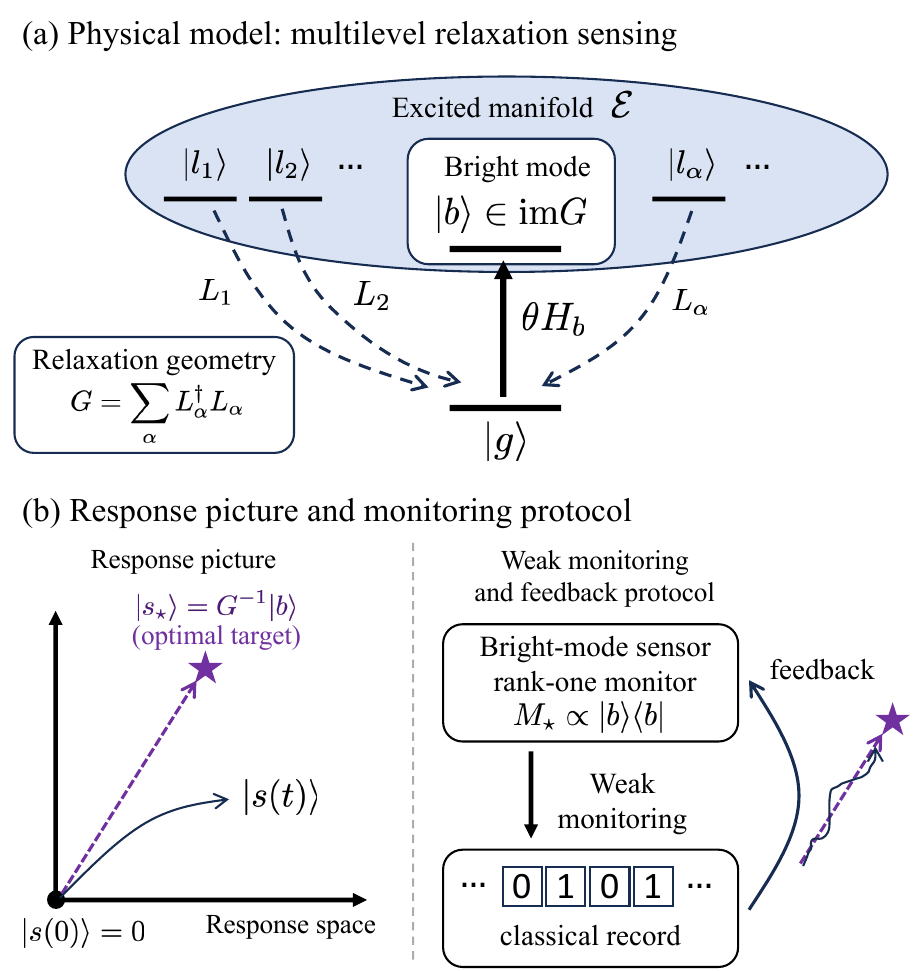}
\caption{Multilevel relaxation sensing and its optimal control. (a) The signal $H_b$ couples the ground state to the excited manifold, which relaxes back to the ground state. (b) Weak monitoring along the bright direction and outcome-dependent feedback stabilize the rate-optimal response $\ket{s_\star}=G^{-1}\ket b$, continuously transferring information to the classical record.}
\label{fig:scheme}
\end{figure}

\paragraph{Multilevel relaxation sensing.---}
We consider a multilevel relaxation sensor consisting of a ground state
$\ket g$ and an excited-state manifold $\mathcal E$, with
$\mathcal H=\operatorname{span}\{\ket g\}\oplus\mathcal E$.
A weak signal with unknown amplitude $\theta$ couples the ground state to the excited manifold through $H_\theta=\theta H_b$, with $H_b=i\left(\ket b\bra g-\ket g\bra b\right)/2$, where $\ket b\in\mathcal E$ is the normalized bright state. The overall coupling strength is absorbed into $\theta$, so $\theta$ has units of frequency and the QFI per unit time has the dimension of time. We focus on the weak-signal regime, i.e., $\theta\approx 0$. We assume a known, time-independent signal coupling and relaxation model. 
The excited manifold relaxes to the ground state through jump operators $L_\alpha=\ket g\bra{\ell_\alpha}$.
The sensor state obeys the master equation~\cite{Gorini1976,Lindblad1976}
\begin{align}\label{eq:master}
    \frac{d\rho_\theta}{dt}=-i[H_\theta,\rho_\theta]+\sum_\alpha \left(L_\alpha\rho_\theta L_\alpha^\dagger-\frac12\{L_\alpha^\dagger L_\alpha,\rho_\theta\}\right).
\end{align}
The vectors $\ket{\ell_\alpha}$ absorb the channel amplitudes and therefore need not be normalized or mutually orthogonal.
The positive operator $G:=\sum_\alpha L_\alpha^\dagger L_\alpha=\sum_\alpha\ket{\ell_\alpha}\bra{\ell_\alpha}$ describes the relaxation geometry within the excited manifold:
its eigenvectors with nonzero eigenvalues are the decay modes, and the corresponding eigenvalues are their decay rates. 
Excited-state superpositions in $\ker G$ are dark to all relaxation channels. 
We assume $\ket b\in\operatorname{im}G$, so the signal couples only to excited components that relax to the ground state. Throughout, $G^{-1}$ denotes the inverse of $G$ on $\operatorname{im}G$. 
The qubit amplitude-damping sensor is the one-dimensional special case. 
Figure~\ref{fig:scheme} summarizes the model and the monitoring protocol developed below.

For the response analysis and the direct-monitoring protocol below, we initialize the sensor in $\rho_\theta(0)=\proj g$ and first consider its unmonitored evolution near $\theta=0$. At $\theta=0$, $H_\theta=0$ and $L_\alpha\ket g=0$; hence, the reference trajectory remains $\rho_0(t)=\proj g$. For $\theta\approx 0$, the state can be expanded as $\rho_\theta(t)=\rho_0(t)+\theta\left.\partial_\theta\rho_\theta(t)\right|_{\theta=0}+O(\theta^2)$. The local distinguishability of the two trajectories is therefore determined by this tangent. Because $H_b$ couples $\ket g$ to the excited manifold, the tangent contains only ground--excited coherence and takes the form~\cite{SM}
\begin{align}
\left.\partial_\theta\rho_\theta\right|_{\theta=0}
&=\frac12\left(\ket s\bra g+\ket g\bra s\right),
\qquad \ket s\in\mathcal E.
\label{eq:vectangent}
\end{align}
The vector $\ket{s(t)}$ thus describes the ground--excited coherence generated by a small change in the signal, while population changes begin at $O(\theta^2)$. It need not be normalized: its components give the response through the different excited-state modes, and the standard QFI formalism applied to the tangent above gives $F_Q(t)=\norm{\ket{s(t)}}^2$ \cite{BraunsteinCaves1994}.

We next determine the time evolution of this response under the competing effects of signal generation and relaxation.
Differentiating the master equation, Eq.~\eqref{eq:master}, at $\theta=0$ gives~\cite{SM}
\begin{align}
\frac{d}{dt}\ket s
&=\ket b-\frac12G\ket s.
\label{eq:vectorfree}
\end{align}
With $\ket{s(0)}=0$, Eq.~\eqref{eq:vectorfree} uniquely determines the free response $\ket{s(t)}$. 
The signal generates ground--excited coherence along $\ket b$, while physical relaxation removes its decay-mode components at generally different rates. 
If $\ket b$ is a decay mode of $G$, the direction of $\ket{s(t)}$ remains fixed and a scalar qubit description suffices. Otherwise the coherence is redistributed among several decay modes as time passes. 
The resulting control problem is to identify the response that maximizes the information-growth rate despite relaxation.

\medskip

\paragraph{Rate-optimal local response.---}
Since $F_Q=\norm{\ket s}^2$, the response dynamics in Eq.~\eqref{eq:vectorfree} directly determine the instantaneous information-growth rate. Rewriting this rate by completing the square gives
\begin{align}
\frac{dF_Q}{dt}&=\bra bG^{-1}\ket b-\norm{G^{1/2}\left(\ket{s(t)}-G^{-1}\ket b\right)}^2.
\label{eq:Rcompletion}
\end{align}
Since the second term is nonnegative, we have $dF_Q/dt\le R$, with equality at $\ket{s(t)}=\ket{s_\star}$.
Therefore, Eq.~\eqref{eq:Rcompletion} gives the rate-optimal response $\ket{s_\star}:=G^{-1}\ket b$ and the maximum information growth rate $R:=\bra bG^{-1}\ket b$. 

In the decay-mode basis, write $G\ket\mu=\gamma_\mu\ket\mu$ with $\gamma_\mu>0$ and $\ket b=\sum_\mu b_\mu\ket\mu$; then $s_{\star,\mu}=b_\mu/\gamma_\mu$ and $R=\sum_\mu|b_\mu|^2/\gamma_\mu$. Long-lived modes therefore support larger responses and contribute more strongly to the attainable rate. 
This same lifetime dependence appears directly in the free decay of the bright state.
Under relaxation alone, its excited-manifold survival probability is $P_{\rm exc}(t)=\bra b e^{-Gt}\ket b$, and hence~\cite{SM}
\begin{align}
\tau_b&:=\int_0^\infty P_{\rm exc}(t)dt=\bra bG^{-1}\ket b=R.
\label{eq:bright_lifetime}
\end{align}
Thus, the ultimate information rate $R$ is exactly the mean population lifetime $\tau_b$ of the bright state. Note that both sides have the dimension of time, since $F_Q$ shares its dimension with $\theta^{-2}$, where $\theta$ has units of frequency.
Importantly, this lifetime is determined by the entire population-decay dynamics, not by the initial decay rate alone. 
Defining the initial decay rate as $\gamma_b=-\dot P_{\rm exc}(0)=\bra bG\ket b$, one has $R\ge 1/\gamma_b$, with equality only when $\ket b$ lies in a single eigenspace of $G$~\cite{SM}.
A single-exponential description based on $\gamma_b$, therefore, underestimates the attainable rate whenever the signal overlaps decay modes with different lifetimes.

Starting from $\ket{s(0)}=0$, each mode with $b_\mu\neq0$ evolves as $s_\mu(t)=2b_\mu(1-e^{-\gamma_\mu t/2})/\gamma_\mu$ and reaches its target component only at $t_\mu=2\ln2/\gamma_\mu$. 
When $\ket b$ overlaps modes with distinct decay rates, free evolution thus cannot reach all target components simultaneously. 
Even for a single decay rate, it cannot maintain the rate-optimal response. 
We therefore seek a control that stabilizes the rate-optimal response $\ket{s_\star}$. 
Below, we derive the unrestricted bound and construct a rank-one monitor that does so asymptotically.

\medskip

\paragraph{Ultimate adaptive benchmark.---}
The benchmark allows unrestricted adaptive
quantum control, including optimization over arbitrary parameter-independent initial states of the sensor and its memories. The sensor may interact with noiseless quantum memories,
undergo intermediate measurements and feedback, and be subjected to arbitrary
parameter-independent controls \cite{PangJordan2017,LiuYuan2017,Poggiali2018}. This class includes ancilla-assisted strategies,
persistent sensor--memory entanglement, and arbitrary QEC encodings. Dividing
the total sensing time $T$ into intervals of duration $dt$, a general
protocol alternates the infinitesimal sensing channel
$\mathcal N_{\theta,dt}$ with control operations on the sensor, auxiliary
memories, and the accumulated classical
record. Taking $dt\to0$ gives the
standard fast-control adaptive setting of Ref.~\cite{Demkowicz2017}. The physical relaxation environment is unobserved; measurements act on the sensor and its auxiliary systems. All QFIs are evaluated at $\theta=0$. We denote
the maximum QFI over this entire class by $F_{\rm opt}(T)$.

The channel-extension (CE) bound and its continuous-time adaptive form apply to this unrestricted class \cite{FujiwaraImai2008,Escher2011,Demkowicz2012,Demkowicz2017,WanLasenby2022}. For the present model, the condition $\ket b\in\operatorname{im}G$ ensures that the signal Hamiltonian can be expressed as a linear combination of the relaxation jump operators and their adjoints. The CE construction then yields a bound linear in the sensing time. Optimizing its coefficients gives the information rate $R=\bra bG^{-1}\ket b$ identified above, as derived in the Supplemental Material~\cite{SM}. Consequently,
\begin{align}
F_{\rm opt}(T)&\le RT\qquad\text{for all }T\ge0.
\label{eq:universal}
\end{align}
The same quantity $R$ bounds both the local QFI growth in Eq.~\eqref{eq:Rcompletion} and the full adaptive problem. The protocol below attains this rate asymptotically.

\medskip

\paragraph{Direct rank-one monitoring.---}
We construct a single monitor acting along the bright direction that attains this bound. The amplitude-damping qubit illustrates the underlying information transfer~\cite{Kurdzialek2025}. In a binary-readout formulation, with qubit decay rate $\gamma$ and scalar response $s$, a weak measurement of strength $\phi$ followed by outcome-dependent feedback leaves $s\cos\phi$ in the sensor and transfers $s^2\sin^2\phi$ of classical Fisher information (FI) to the record. Taking $\sin^2\phi=2m dt+o(dt)$ gives $\dot s=1-(\gamma/2+m)s$ and $\dot F_{\rm rec}=2ms^2$, where $F_{\rm rec}$ is the accumulated FI in the measurement record, $m$ is the monitoring rate, and feedback realigns the two conditional branches. 

For the multilevel sensor, the rate-optimal response $\ket{s_\star}=G^{-1}\ket b$ generally spans several decay modes, but monitoring along the bright direction stabilizes the full response. Using the bright state $\ket b$ defined above, choose
\begin{align}
m_\star&:=\frac{1}{2R},&
M_\star&:=m_\star\proj b=\frac{\ket b\bra b}{2R}.
\label{eq:Mstar}
\end{align}
Because $m_\star=1/(2\tau_b)$, the population-decay curve determines both the ultimate rate and the monitor strength. 
The derivation from a general rank-one monitor is given in the Supplemental Material~\cite{SM}. After each weak readout, the outcome is stored in the classical record, while an outcome-dependent, parameter-independent feedback unitary realigns the conditional reference state and local response. The resulting corrected dynamics are
\begin{align}
\frac{d}{dt}\ket s&=\ket b-\frac12G\ket s-M_\star\ket s,&
\dot F_{\rm rec}&=2\bra sM_\star\ket s.
\label{eq:directdyn}
\end{align}
The first equation tracks the sensitivity still carried by the sensor, whose squared norm is the residual sensor QFI, while the second tracks the FI already transferred to the record. At any chosen stopping time $T$, an optimal final sensor readout therefore gives
$F_{\rm dir}(T)=F_{\rm rec}(T)+\norm{\ket{s(T)}}^2$~\cite{SM}.

Differentiating $\norm{\ket s}^2$ with Eq.~\eqref{eq:directdyn} and adding the record-FI rate yields the information balance
\begin{align}
\frac{d}{dt}\left(F_{\rm rec}+\norm{\ket s}^2\right)
&=R-\norm{G^{1/2}(\ket s-\ket{s_\star})}^2.
\label{eq:ledger}
\end{align}
The monitoring terms cancel: weak measurement transfers sensitivity from the sensor to the record, while feedback realigns the conditional branches and allows this transfer to be repeated. The relaxation-weighted mismatch from $\ket{s_\star}$ determines the nonnegative deficit from $R$. Moreover, $M_\star\ket{s_\star}=\ket b/2$ and $G\ket{s_\star}=\ket b$; hence Eq.~\eqref{eq:directdyn} has $\ket{s_\star}$ as a steady response. At this operating point, the sensor QFI remains constant and the record FI grows at rate $R$. No direct control of the other decay modes is required. Related outcome-resolved information-conservation relations for weak measurement and reversal in decaying multilevel systems were derived in Ref.~\cite{Maleki2026}; here the information balance is tied directly to the ultimate information rate and to a rank-one monitor that attains it.

The monitor can be implemented as follows. Define $X_b:=\ket b\bra g+\ket g\bra b$ and $Y_b:=-i\ket g\bra b+i\ket b\bra g$. Each fresh meter performs the binary weak measurement $E_r=[I+r\sin\phi X_b]/2$ for $r=\pm1$, with $K_r=\sqrt{E_r}$, followed by $V_r=e^{ir\phi Y_b/2}$. Taking $\sin^2\phi=2m_\star dt+o(dt)$ realizes $M_\star$: the readout transfers sensitivity along the bright direction to the record at the rate in Eq.~\eqref{eq:directdyn}, while $V_r$ realigns the two branches. The explicit Kraus calculation is given in the Supplemental Material~\cite{SM}. Each meter is measured and discarded, and only its classical outcome persists.

To quantify the finite-time approach to the steady rate, start from the ground state, $\ket{s(0)}=0$, and define $\ket{\delta(t)}=\ket{s(t)}-\ket{s_\star}$. Integrating Eq.~\eqref{eq:ledger} gives
\begin{align}
F_{\rm rec}(T)+\norm{\ket{s(T)}}^2
&=RT-\int_0^T\norm{G^{1/2}\ket{\delta(t)}}^2dt.
\label{eq:integrated}
\end{align}
Thus, the deficit from $RT$ is the accumulated mismatch as the sensor response approaches $\ket{s_\star}$. Let $A=G/2+M_\star$. Since $M_\star\ket{s_\star}=\ket b/2$, we have $\dot{\ket\delta}=-A\ket\delta$, whose squared-norm decay bounds the accumulated mismatch by the $T$-independent constant $C:=\norm{\ket{s_\star}}^2$~\cite{SM}.

Combining this finite transient bound with Eq.~\eqref{eq:integrated} and the unrestricted bound in Eq.~\eqref{eq:universal} yields
\begin{align}
RT-C
&\le F_{\rm dir}(T)
\le F_{\rm opt}(T)
\le RT.
\label{eq:finalsandwich}
\end{align}
For every sensing time, the direct-monitoring protocol remains within the fixed additive gap $C$ of the unrestricted optimum. Since $C$ does not grow with $T$, the difference becomes asymptotically negligible, and both strategies attain the same information rate,
\begin{align}
\lim_{T\to\infty}\frac{F_{\rm dir}(T)}{T}
&=\lim_{T\to\infty}\frac{F_{\rm opt}(T)}{T}
=R=\tau_b.
\label{eq:rate}
\end{align}

\begin{figure}[tb]
\centering
\includegraphics[width=\columnwidth]{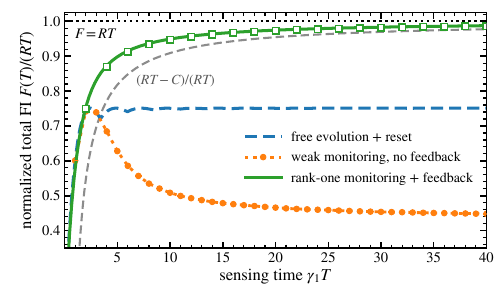}
\caption{Normalized total FI $F(T)/(RT)$ for $G=\gamma_1\operatorname{diag}(1,9)$ in the decay-mode basis $\{\ket{\mu_1},\ket{\mu_2}\}$ and $\ket b=(\ket{\mu_1}+\ket{\mu_2})/\sqrt2$. At each sensing time $T$, the free-evolution interrogation period and the no-feedback monitoring rate are separately optimized to maximize the total FI. Open squares show the finite-meter implementation of rank-one monitoring with feedback. Error bars on the no-feedback circles are one Monte Carlo standard error and are smaller than the symbols. The reference curves are the ultimate bound $F=RT$ and the finite-time lower bound $F=RT-C$. Numerical details are given in the Supplemental Material~\cite{SM}.}
\label{fig:numerics}
\end{figure}

\paragraph{Numerical illustration.---}
Figure~\ref{fig:numerics} illustrates the analytical predictions for two unequal decay modes. For this example, $R=5/(9\gamma_1)$ and $\ket{s_\star}=G^{-1}\ket b$ is not parallel to $\ket b$, whereas a single-exponential approximation based on the initial decay rate gives only $1/\gamma_b=1/(5\gamma_1)$. The direct-monitoring protocol approaches $R$ while satisfying the finite-time guarantee $RT-C$, and its finite-meter implementation closely follows the continuous-monitoring result. Free evolution with periodic readout and reset remains below $R$ because it cannot sustain the rate-optimal response, while monitoring without feedback exhibits branch-dependent sensitivity drift. Monitoring with feedback removes both limitations, stabilizes the rate-optimal response, and attains the ultimate rate.

\medskip

\paragraph{Entanglement cannot increase the rate.---}
Consider $n$ identical sensors with independent local relaxation
channels, signal Hamiltonian
$H_\theta^{(n)}=\theta\sum_{j=1}^n H_b^{(j)}$, and jump operators
$L_{j\alpha}=L_\alpha^{(j)}$. We allow arbitrary entangled initial states,
intersensor controls, collective measurements, noiseless quantum memories,
feedback, and joint QEC, and denote the optimum over this full joint adaptive
class by $F_{\rm opt}^{(n)}(T)$.

Over time $T$, these sensors use the same infinitesimal channel $nT/dt$ times. The adaptive channel-extension bound includes parallel entangled strategies and collective controls; applying the same single-sensor CE construction to this total number of channel uses gives $F_{\rm opt}^{(n)}(T)\le nRT$ in the continuous-time limit~\cite{Demkowicz2017,SM}. Conversely, applying the single-sensor rank-one monitoring protocol independently to each sensor, with local readouts and feedback, gives
$F_{\rm dir}^{(n)}(T)=nF_{\rm dir}(T)$ and therefore
\begin{align}
nRT-nC
&\le F_{\rm dir}^{(n)}(T)
\le F_{\rm opt}^{(n)}(T)
\le nRT.
\label{eq:manysensor}
\end{align}
Dividing by $nT$ and taking $T\to\infty$ shows that both the unrestricted
optimum and the product direct-monitoring protocol attain the rate $R$ per
sensor. Thus, arbitrary intersensor entanglement and joint QEC cannot increase
the asymptotic information rate per sensor. They may modify finite-time
performance, but their advantage over product direct monitoring is bounded by
the time-independent quantity $nC$. This conclusion relies on independent
local relaxation; collective relaxation can produce a different joint rate.

\medskip

In summary, the mean population lifetime of the bright state is the ultimate information rate for sensing a weak signal under multilevel relaxation. The relaxation operator $G$ also determines the rate-optimal response $\ket{s_\star}=G^{-1}\ket b$. Although this response generally spans several decay modes, monitoring along the bright direction maintains it: weak measurements transfer information to the record, and classical feedback realigns the conditional branches. The information deficit remains bounded in time, so the protocol reaches the ultimate information rate asymptotically. Applied independently to $n$ sensors, it attains $nR$, which cannot be improved by intersensor entanglement or joint QEC. It remains to determine how these results change under more general dissipation, including multiple decay sinks, dephasing, and thermal excitation, and under finite measurement efficiency~\cite{Len2022}, feedback delay, and control bandwidth.

\medskip

\begin{acknowledgments}
The authors used ChatGPT (OpenAI; GPT-5.6 Sol and GPT-6 Astra) to assist with analytical cross-checks, numerical simulations, and manuscript editing. All outputs were independently verified by the authors.
C.O., S.H.L., and Y.L. were supported by the Institute of Information \& Communications Technology Planning \& Evaluation (IITP) grant funded by the Korean government (Ministry of Science and ICT (MSIT)) (No. IITP-2025-RS-2025-02283189).
C.O. and Y.L. were supported by the National Research Foundation of Korea (NRF) grant funded by the Korean government (MSIT) (No. RS-2024-00431768).
C.O. was supported by the NRF grant (No. RS-2025-00515456) and the IITP grants (No. RS-2024-00437284 and No. IITP-2025-RS-2025-02263264), funded by the Korean government (MSIT).
S.H.L. was supported by the start-up fund and the 2026 Research Fund (1.250007.01, 1.260015.01) of Ulsan National Institute of Science \& Technology (UNIST) and the NRF grants (No. RS-2025-25464492 and No. RS-2026-25520808).
\end{acknowledgments}

\bibliography{reference}

\end{document}


\definecolor{navy}{RGB}{46,72,102}
\definecolor{pink}{RGB}{219,48,122}
\definecolor{grey}{RGB}{184,184,184}
\definecolor{yellow}{RGB}{255,192,0}
\definecolor{grey1}{RGB}{217,217,217}
\definecolor{grey2}{RGB}{166,166,166}
\definecolor{grey3}{RGB}{89,89,89}
\definecolor{red}{RGB}{255,0,0}

\title{Supplemental Material for ``Ultimate Information Rate for Quantum Sensing under Multilevel Relaxation''}
\author{Changhun Oh}
\email{changhun0218@gmail.com}
\affiliation{Department of Physics, Korea Advanced Institute of Science and Technology, Daejeon 34141, Korea}
\author{Seok Hyung Lie}
\email{seokhyung@unist.ac.kr}
\affiliation{Department of Physics, Ulsan National Institute of Science and Technology, Ulsan 44919, Korea}
\author{Youngrong Lim}
\email{sshaep@gmail.com}
\affiliation{Department of Physics, Chungbuk National University, Cheongju 28644, Korea}

\maketitle

\section{Multilevel relaxation tangent and inverse-relaxation geometry}
\label{sec:SM_multilevel}

This section derives the linear-response dynamics generated by multilevel relaxation and identifies the inverse-relaxation geometry controlling local quantum Fisher information (QFI) growth. It establishes the response vector $\ket s$, the rate-optimal response $\ket{s_\star}=G^{-1}\ket b$, and the rate $R=\bra bG^{-1}\ket b$ used throughout the remaining sections. Here, $G^{-1}$ denotes the inverse of $G$ on $\operatorname{im}G$, extended by zero on $\ker G$ when acting on the full excited manifold. As in the main text, we absorb the overall coupling strength into $\theta$ and normalize the bright state, $\langle b|b\rangle=1$. Thus, $\theta$ is the effective coupling strength between the ground state and the bright state. All QFIs and FI rates below are local at $\theta=0$, with fixed, known coupling direction and relaxation operators. The sensing model is
\begin{align}
H_b&=\frac{i}{2}(\ket b\bra g-\ket g\bra b),\qquad
L_\alpha=\ket g\bra{\ell_\alpha},\qquad
G=\sum_\alpha\ket{\ell_\alpha}\bra{\ell_\alpha},\qquad \ket b\in\operatorname{im}G.
\label{eq:SM_multilevel_model_compact}
\end{align}
For the local-response analysis in this section, initialize the sensor in the parameter-independent state $\rho_\theta(0)=\proj g$. The unrestricted benchmark in Sec.~\ref{sec:SM_adaptive} allows arbitrary parameter-independent initial states. Write $\rho_\theta(t)=\rho_0(t)+\theta\left.\partial_\theta\rho_\theta(t)\right|_{\theta=0}+O(\theta^2)$. The state obeys the master equation~\cite{Gorini1976,Lindblad1976}
\begin{align}
    \frac{d\rho_\theta}{dt}=-i[\theta H_b,\rho_\theta]+\sum_\alpha\mathcal D[L_\alpha](\rho_\theta),\qquad\mathcal D[L](\rho):=L\rho L^\dagger-\frac12\{L^\dagger L,\rho\}.
\end{align}

To obtain the first-order dynamics, define $\sigma(t):=\left.\partial_\theta\rho_\theta(t)\right|_{\theta=0}$. Because the jump operators are parameter independent, differentiation acts only on the state in the dissipative terms, while the Hamiltonian term produces the source.

Explicitly,
\begin{align}
\left.\partial_\theta\bigl[-i\theta[H_b,\rho_\theta]\bigr]\right|_{\theta=0}&=-i[H_b,\rho_0],\qquad
\left.\partial_\theta\mathcal D[L_\alpha](\rho_\theta)\right|_{\theta=0}=\mathcal D[L_\alpha](\sigma).
\label{eq:SM_first_order_terms}
\end{align}

Therefore, the first-order tangent obeys
\begin{align}
\frac{d\sigma}{dt}=-i[H_b,\rho_0]+\sum_\alpha\mathcal D[L_\alpha](\sigma).
\end{align}

Here, the reference state is stationary: $H_{\theta=0}=0$ and $L_\alpha\ket g=0$ imply $\rho_0(t)=\ket g\bra g$. The source in the tangent equation is therefore
$-i[H_b,\rho_0]=(\ket b\bra g+\ket g\bra b)/2$: it contains only Hermitian coherences between the ground state and the excited manifold, with no population or excited--excited component. This motivates the coherence space
\begin{align}
\mathcal C_{ge}:=\left\{Q_v=\frac12(\ket v\bra g+\ket g\bra v):\ket v\in\mathcal E\right\}.
\end{align}
Here, $\ket v$ is an arbitrary complex vector. Its real and imaginary components encode the two ground--excited coherence quadratures, so $\mathcal C_{ge}$ contains the full Hermitian tangent sector generated by the signal.

The first-order equation is closed on this space. The parameter-independent initial state gives $\sigma(0)=0\in\mathcal C_{ge}$, and the Hamiltonian source is $Q_b\in\mathcal C_{ge}$. For every $Q_v\in\mathcal C_{ge}$,
\begin{align}
L_\alpha Q_vL_\alpha^\dagger&=0,\qquad
\sum_\alpha\mathcal D[L_\alpha](Q_v)
=-\frac14(G\ket v\bra g+\ket g\bra vG)
=Q_{-Gv/2}\in\mathcal C_{ge}.
\end{align}
The jump term vanishes, while the anticommutator maps the excited-space vector $\ket v$ to $-G\ket v/2$ without generating any other operator block. Thus, the initial condition, the source, and the relaxation dynamics all remain in $\mathcal C_{ge}$. It follows that $\sigma(t)=Q_{s(t)}$ for a response vector $\ket{s(t)}\in\mathcal E$. Substitution into the first-order equation gives
\begin{align}
\left.\partial_\theta\rho_\theta(t)\right|_{\theta=0}&=\frac12\left(\ket{s(t)}\bra g+\ket g\bra{s(t)}\right),\qquad
\frac{d}{dt}\ket s=\ket b-\frac12G\ket s.
\label{eq:SM_vector_free_compact}
\end{align}
Here, $\ket s$ is a response vector and need not be normalized.

The symmetric logarithmic derivative (SLD) is $L_0=\ket s\bra g+\ket g\bra s$, so the QFI stored in the sensor is~\cite{BraunsteinCaves1994}
\begin{align}
F_Q(t)=\Tr[\rho_0L_0^2]=\langle s(t)|s(t)\rangle.
\label{eq:SM_vector_qfi_compact}
\end{align}

Differentiating the sensor QFI and using Eq.~\eqref{eq:SM_vector_free_compact} gives
\begin{align}
\frac{d}{dt}\norm{\ket s}^2
=2\operatorname{Re}\langle s|b\rangle-\langle s|G|s\rangle 
=\bra bG^{-1}\ket b-\norm{G^{1/2}(\ket s-G^{-1}\ket b)}^2.
\label{eq:SM_information_geometry_compact}
\end{align}
The second equality follows by completing the square on $\operatorname{im}G$. Since $\ket b\in\operatorname{im}G$, we have $GG^{-1}\ket b=\ket b$. Expanding $\norm{G^{1/2}(\ket s-G^{-1}\ket b)}^2$ then gives $\langle s|G|s\rangle-2\operatorname{Re}\langle s|b\rangle+\bra bG^{-1}\ket b$, which proves the identity.
Equation~\eqref{eq:SM_information_geometry_compact} then identifies the rate-optimal response and its maximal instantaneous QFI-growth rate as
\begin{align}
\ket{s_\star}&:=G^{-1}\ket b,\qquad
R:=\bra bG^{-1}\ket b.
\label{eq:SM_geometry_compact}
\end{align}

\medskip
The information rate $R$ can also be read from a free population-decay experiment. Prepare the normalized bright state $\ket b$ and switch off the signal and controls. Let $\varrho(t)$ be the sensor state, let $P_{\mathcal E}$ project onto the excited manifold, and write $\varrho_{\mathcal E}(t):=P_{\mathcal E}\varrho(t)P_{\mathcal E}$. Each jump transfers population to $\ket g$, so the projected master equation and its solution are
\begin{align}
\frac{d\varrho_{\mathcal E}}{dt}&=-\frac12\{G,\varrho_{\mathcal E}\},\qquad
\varrho_{\mathcal E}(0)=\proj b,\qquad
\varrho_{\mathcal E}(t)=e^{-Gt/2}\proj b e^{-Gt/2}.
\label{eq:SM_bright_population}
\end{align}
Consequently, the probability that the excitation has not yet decayed is $P_{\rm exc}(t)=\Tr\varrho_{\mathcal E}(t)=\bra b e^{-Gt}\ket b$. Integrating this survival probability gives the mean population lifetime. Writing $G\ket\mu=\gamma_\mu\ket\mu$ for the decay modes and using $\ket b\in\operatorname{im}G$, so that only modes with $\gamma_\mu>0$ contribute, we obtain
\begin{align}
\tau_b&:=\int_0^\infty P_{\rm exc}(t)dt
=\sum_{\mu:\gamma_\mu>0}\frac{|\langle\mu|b\rangle|^2}{\gamma_\mu}
=\bra bG^{-1}\ket b,\qquad
R=\tau_b.
\label{eq:SM_bright_lifetime}
\end{align}
The channel-extension (CE) bound and the attainable protocol derived below establish that this lifetime is the optimal asymptotic information rate.

To distinguish this lifetime from the inverse initial decay rate, define $\gamma_b:=-\dot P_{\rm exc}(0)=\bra bG\ket b$. Cauchy--Schwarz on $\operatorname{im}G$ gives
\begin{align}
1=|\bra bG^{1/2}G^{-1/2}\ket b|^2
&\le\bra bG\ket b\bra bG^{-1}\ket b=\gamma_bR,\qquad R\ge\gamma_b^{-1}.
\label{eq:SM_initial_decay_comparison}
\end{align}
Equality holds exactly when $G^{1/2}\ket b$ and $G^{-1/2}\ket b$ are parallel, equivalently when $G\ket b=\gamma_b\ket b$. In that case, $P_{\rm exc}(t)=e^{-\gamma_b t}$ and the ground--bright qubit sector is closed under relaxation. If the signal overlaps distinct decay rates, the inequality is strict: averaging the rates before taking the inverse loses the lifetime information relevant to sensing. For $G=\gamma_1\operatorname{diag}(1,9)$ and an equally weighted bright state, $R=5/(9\gamma_1)$, whereas the single-exponential estimate gives $\gamma_b^{-1}=1/(5\gamma_1)$.

\section{Unrestricted adaptive benchmark and channel-extension bound}
\label{sec:SM_adaptive}

This section defines the unrestricted adaptive strategy class and derives the CE upper bound for the multilevel relaxation model. It connects the general fast-control benchmark to the information rate $R$ identified in Sec.~\ref{sec:SM_multilevel}. The sensor may be coupled to arbitrary noiseless quantum memories and classical registers, with parameter-independent control applied throughout the sensing evolution. This framework includes coherent sensor-memory interactions, intermediate measurements, feedback, resets, and quantum error correction (QEC). The physical relaxation environment remains unobserved.

Let $S$ denote the sensor, $A$ a noiseless quantum memory, and $C$ a classical register containing the accumulated measurement record. Divide the total sensing time $T$ into $N$ intervals of duration $dt=T/N$. A general adaptive protocol can be written recursively as
\begin{align}
\Omega_{\theta,j}
&=
\mathcal C_j
\circ
\left(
\mathcal N_{\theta,dt}
\otimes
\mathcal I_{AC}
\right)
(\Omega_{\theta,j-1}),
\qquad
j=1,\ldots,N,
\label{eq:SM_adaptive_recursion}
\end{align}
where $\Omega_0$ is an arbitrary parameter-independent initial state of $SAC$, including sensor--memory entanglement:
\begin{align}
\Omega_{\theta,0}
&=
\Omega_0.
\label{eq:SM_adaptive_initial}
\end{align}
Here, $\mathcal N_{\theta,dt}$ is the physical sensing channel acting on the sensor during one interval, $\mathcal I_{AC}$ is the identity channel on $AC$, and each $\mathcal C_j$ is a parameter-independent completely positive trace-preserving (CPTP) map acting jointly on $SAC$.

The quantum memory $A$ may remain coherently coupled to the sensor throughout the protocol. Intermediate measurement outcomes can be stored in orthogonal states of $C$, allowing subsequent controls to depend on the complete measurement history. The maps $\mathcal C_j$ may also implement state preparation, resets, syndrome extraction, conditional recovery, and arbitrary QEC or approximate-QEC cycles. Thus, the recursion above represents the full sensor-side adaptive control class considered in the main text.

At the end of the protocol, a measurement history $h$ naturally gives a classical-quantum state of the form
\begin{align}
\Omega_\theta
&=
\sum_h
p_h(\theta)
\ket h\bra h_C
\otimes
\rho_{h,\theta}.
\label{eq:SM_cq_state}
\end{align}
Its QFI separates into the classical Fisher information (FI) contained in the record and the average QFI retained in the conditional quantum state~\cite{Gammelmark2014,Albarelli2018},
\begin{align}
F_Q(\Omega_\theta)
&=
F_C[\{p_h\}]
+
\sum_h
p_h(\theta)
F_Q(\rho_{h,\theta}).
\label{eq:SM_cq_QFI}
\end{align}
Here, $F_C[\{p_h\}]$ denotes the classical FI of the outcome distribution $\{p_h(\theta)\}$. This decomposition provides the information accounting used throughout the direct-monitoring analysis: information already transferred to the classical record and information still stored in the conditional sensor-memory state are both included in the total metrological performance.

Taking $N\to\infty$ gives the fast-control adaptive setting of Ref.~\cite{Demkowicz2017}, with arbitrarily frequent parameter-independent controls during a fixed total sensing time $T$. We define $F_{\rm opt}(T)$ as the maximum QFI attainable over this class. The direct-monitoring protocol introduced below is a particular strategy within the same benchmark.

\subsection{Channel-extension bound for adaptive Markovian metrology}
\label{sec:SM_CE_recall}

We summarize the standard CE construction of Ref.~\cite{Demkowicz2017} through Eq.~\eqref{eq:SM_CE_linear_optimization}, retaining the steps needed for the model-specific minimization below. A complementary continuous-time treatment is given in Ref.~\cite{WanLasenby2022}. Write a sensing channel as $\mathcal N_\theta(\rho)=\sum_a K_{a,\theta}\rho K_{a,\theta}^\dagger$ and collect them into the operator-valued column $\bm K_\theta=(K_{0,\theta},K_{1,\theta},\ldots)^{\T}$, whose entries act on the sensor Hilbert space.
After padding with zero operators when necessary, equivalent Kraus representations of the same physical channel are related by a $\theta$-dependent unitary $u_\theta$, $\widetilde{\bm K}_\theta=u_\theta\bm K_\theta$.
Componentwise, the transformation and the invariance of the channel read
\begin{align}
\widetilde K_{a,\theta}&=\sum_b(u_\theta)_{ab}K_{b,\theta},\qquad
\sum_a\widetilde K_{a,\theta}\rho\widetilde K_{a,\theta}^\dagger=\sum_aK_{a,\theta}\rho K_{a,\theta}^\dagger.
\label{eq:SM_Kraus_gauge_equivalence}
\end{align}
We choose $u_{\theta=0}=I$ without loss of generality. Differentiating the unitarity condition at $\theta=0$, we write $\left.\partial_\theta u_\theta\right|_{\theta=0}=-i\mathsf h$, where $\mathsf h=\mathsf h^\dagger$.
Thus, $\bm K_\theta$ is the original Kraus column and $\widetilde{\bm K}_\theta$ is an equivalent Kraus column for the same channel. At the operating point, let $\bm K:=\bm K_{\theta=0}$ and $\bm K':=\left.\partial_\theta\bm K_\theta\right|_{\theta=0}$. Since $u_0=I$, one has $\widetilde{\bm K}=\bm K$, while differentiating $\widetilde{\bm K}_\theta=u_\theta\bm K_\theta$ gives the first relation below. We define the two sensor-space operators entering the CE bound by
\begin{align}
\widetilde{\bm K}'&=\bm K'-i\mathsf h\bm K,\qquad
\alpha:=\widetilde{\bm K}'^\dagger\widetilde{\bm K}',\qquad
\beta:=-i\widetilde{\bm K}'^\dagger\widetilde{\bm K}.
\label{eq:SM_CE_alpha_beta}
\end{align}
The positive operator $\alpha$ is the sum of squared parameter derivatives over the Kraus operators for one channel use, whereas $\beta$ is the derivative--Kraus overlap operator whose contributions can add coherently between different uses. Here, $\norm{\cdot}$ denotes the operator norm on the sensor Hilbert space. For $N$ channel uses interspersed with arbitrary parameter-independent controls, the CE bound reads~\cite{Demkowicz2017}
\begin{align}
F_Q^{(N)}
\le 4\min_{\widetilde{\bm K}_\theta,x>0}
\left\{N\norm{\alpha}+N(N-1)\norm{\beta}
\left(x\norm{\alpha}+\norm{\beta}+\frac1x\right)\right\}.
\label{eq:SM_CE_discrete}
\end{align}

The minimization over $\widetilde{\bm K}_\theta$ uses the unitary freedom $\widetilde{\bm K}_\theta=u_\theta\bm K_\theta$ to make the upper bound as small as possible. 
We therefore seek a generator $\mathsf h$ that removes the terms amplified by powers of $N$ in the continuous-time limit.

For the Markovian generator $\dot\rho_\theta=-i\theta[H,\rho_\theta]+\sum_j\mathcal D[L_j](\rho_\theta)$, the corresponding infinitesimal Kraus representation is~\cite{Demkowicz2017}
\begin{align}
K_{0,\theta}&=I-\left(\frac12\sum_jL_j^\dagger L_j+i\theta H\right)dt+O(dt^2),\qquad
K_{j,\theta}=L_j\sqrt{dt}+O(dt^{3/2}).
\label{eq:SM_CE_general_Kraus}
\end{align}
Before optimizing over equivalent Kraus representations, first take $\widetilde{\bm K}_\theta=\bm K_\theta$, i.e., $u_\theta=I$ and $\mathsf h=0$. At $dt=0$, $\bm K=(I,0,\ldots)^{\T}$ and $\bm K'=0$. For this choice, Eq.~\eqref{eq:SM_CE_alpha_beta} gives $\alpha=H^2dt^2+O(dt^3)$ and $\beta=Hdt+O(dt^2)$.

For a general $dt$-dependent unitary $u_\theta$, decompose its Hermitian generator in the basis indexed by the no-jump and jump Kraus operators:
\begin{align}
\mathsf h(dt)&=
\begin{pmatrix}
h_{00}&\bm r^\dagger\\
\bm r&\mathsf c
\end{pmatrix},
\qquad h_{00}\in\mathbb R,\quad \mathsf c=\mathsf c^\dagger.
\label{eq:SM_h_general_blocks}
\end{align}
For $J$ jump operators, $\bm r=(r_1,\ldots,r_J)^{\T}$ and $\mathsf c=(c_{jk})$ may depend on $dt$. Using $K_0=I+O(dt)$, $K_j=L_j\sqrt{dt}+O(dt^{3/2})$, $K_0'=-iHdt+O(dt^2)$, and $K_j'=O(dt^{3/2})$, define
\begin{align}
A_0&:=h_{00}I+\sqrt{dt}\bm r^\dagger\bm L+Hdt,\qquad
A_j:=r_jI+\sqrt{dt}\sum_kc_{jk}L_k,
\label{eq:SM_gauge_derivative_blocks}
\end{align}
where $\bm L=(L_1,\ldots,L_J)^{\T}$. The transformed derivatives are $\widetilde K_0'=-iA_0+\cdots$ and $\widetilde K_j'=-iA_j+\cdots$. Substitution in Eq.~\eqref{eq:SM_CE_alpha_beta} displays the leading contributions to both CE operators:
\begin{align}
\alpha=A_0^\dagger A_0+\sum_jA_j^\dagger A_j+\cdots,\qquad
\beta=Hdt+h_{00}I+\sqrt{dt}\left(\bm r^\dagger\bm L+\bm L^\dagger\bm r\right)+dt\bm L^\dagger\mathsf c\bm L+\cdots.
\label{eq:SM_alpha_beta_power_counting}
\end{align}
The omitted terms are of higher order after imposing the scalings obtained below. Equation~\eqref{eq:SM_alpha_beta_power_counting} shows directly that an $O(1)$ contribution to $h_{00}$ produces $O(1)$ terms in both $\alpha$ and $\beta$. Likewise, an $O(1)$ contribution to $\bm r$ makes $\sum_jA_j^\dagger A_j=O(1)$ and contributes to $\beta$ at order $\sqrt{dt}$. The jump--jump block $\mathsf c$, however, always appears multiplied by $\sqrt{dt}$ in $A_j$ and by $dt$ in $\beta$, so it may remain $O(1)$. These contributions motivate the following sufficient scaling family for keeping $\alpha,\beta=O(dt)$:
\begin{align}
h_{00}&=O(dt),\qquad
\bm r=O(\sqrt{dt}),\qquad
\mathsf c=O(1).
\label{eq:SM_h_scaling}
\end{align}
Write $h_{00}=a(dt)dt$ and $\bm r=\bm\eta(dt)\sqrt{dt}$. We use the regular coefficient expansions
\begin{align}
a(dt)&=a+O(dt),\qquad
\bm\eta(dt)=\bm\eta+O(dt),\qquad
\mathsf c(dt)=\mathsf c+O(dt),
\label{eq:SM_regular_coefficients}
\end{align}
which give $\alpha=\alpha^{(1)}dt+O(dt^2)$ and $\beta=\beta^{(1)}dt+O(dt^2)$. This explicit family is sufficient for the tight bound derived below.

The CE bound contains the $x$-independent term $N(N-1)\norm{\beta}^2$, which carries a quadratic factor in the number of channel uses. Whether its leading contribution can be canceled depends on the physical model. For a choice with $\beta^{(1)}\ne0$, taking $N=T/dt$ retains $T^2\norm{\beta^{(1)}}^2$. This term alone prevents that choice from yielding the desired linear bound; the $x$-dependent terms must also be controlled to obtain any finite continuous-time bound. Within the regular family above, cancellation of $\beta^{(1)}$ gives $\beta=O(dt^2)$. Setting $x=dt^{-1/2}$ then makes all $\beta$-dependent terms vanish at fixed $T$: the $\norm{\beta}^2$ contribution is $O(T^2dt^2)$ and the remaining contributions are $O(T^2\sqrt{dt})$. Consequently, minimizing over choices in this regular family with $\beta^{(1)}=0$ gives
\begin{align}
F_Q(T)&\le4T\min_{\mathsf h:\beta^{(1)}=0}\norm{\alpha^{(1)}}.
\label{eq:SM_CE_linear_optimization}
\end{align}
In the following subsection, we show that $\ket b\in\operatorname{im}G$ guarantees such a choice for multilevel relaxation and construct the required $\mathsf h(dt)$ explicitly.

\subsection{Application to the multilevel relaxation model}
\label{sec:SM_CE_relaxation}

We seek a Kraus-gauge generator satisfying $\beta^{(1)}=0$. This removes the leading contribution multiplied by $N(N-1)$ in Eq.~\eqref{eq:SM_CE_discrete} and leaves the linear-in-$T$ bound in Eq.~\eqref{eq:SM_CE_linear_optimization}. The condition $\ket b\in\operatorname{im}G$ will guarantee that such a generator exists. For $H=H_b$ and $L_j=\ket g\bra{\ell_j}$, Eq.~\eqref{eq:SM_CE_general_Kraus} gives
\begin{align}
K_0&=I-\frac12Gdt+O(dt^2),\qquad K_j=L_j\sqrt{dt}+O(dt^{3/2}),\qquad K_0'=-iH_bdt+O(dt^2),\qquad K_j'=O(dt^{3/2}).
\label{eq:SM_model_Kraus_orders}
\end{align} 
Let $a$, $\bm\eta$, and $\mathsf c$ denote the finite leading coefficients introduced after Eq.~\eqref{eq:SM_h_scaling}:
\begin{align}
h_{00}&=a dt+O(dt^2),\qquad h_{j0}=\eta_j\sqrt{dt}+O(dt^{3/2}),\qquad h_{jk}=c_{jk}+O(dt),
\qquad a\in\mathbb R,\quad \mathsf c=\mathsf c^\dagger.
\label{eq:SM_model_h_general}
\end{align}
Here, $\bm\eta=(\eta_1,\ldots,\eta_J)^{\T}\in\mathbb C^J$, and its component $\eta_j$ is associated with the jump Kraus operator $K_{j,\theta}$. Writing $\bm L=(L_1,\ldots,L_J)^{\T}$ and $B_j:=\eta_j I+\sum_k c_{jk}L_k$, substitution into $\widetilde K_a'=K_a'-i\sum_bh_{ab}K_b$ gives
\begin{align}
\widetilde K_0'
&=-i\left(H_b+aI+\bm\eta^\dagger\bm L\right)dt+O(dt^2),\qquad
\widetilde K_j'=-iB_j\sqrt{dt}+O(dt^{3/2}).
\label{eq:SM_gauged_derivatives_general}
\end{align}
Substitution into Eq.~\eqref{eq:SM_CE_alpha_beta} gives the leading coefficients
\begin{align}
\alpha^{(1)}=\sum_j B_j^\dagger B_j,\qquad
\beta^{(1)}=H_b+aI+\bm\eta^\dagger\bm L+\bm L^\dagger\bm\eta+\bm L^\dagger\mathsf c\bm L.
\label{eq:SM_CE_general_first_order}
\end{align}
The operator blocks separate according to their action on $\ket g$ and the excited manifold. The term $H_b$ lies entirely in the ground--excited block. The scalar $aI$ is the only contribution to the ground--ground block, so $\beta^{(1)}=0$ requires $a=0$. The term $\bm L^\dagger\mathsf c\bm L$ acts only within the excited manifold and is unnecessary for canceling $H_b$; we may set $\mathsf c=0$. The remaining terms $\bm\eta^\dagger\bm L+\bm L^\dagger\bm\eta$ have exactly the ground--excited structure of $H_b$. Moreover, $\ket b\in\operatorname{im}G=\operatorname{span}\{\ket{\ell_j}\}$ ensures that a suitable $\bm\eta$ exists.

It is therefore sufficient to use the generator
\begin{align}
\mathsf h(dt)&=\sqrt{dt}
\begin{pmatrix}0&\bm\eta^\dagger\\ \bm\eta&0_{J\times J}\end{pmatrix}.
\label{eq:SM_gauged_derivatives}
\end{align}
For this choice, Eq.~\eqref{eq:SM_CE_general_first_order} reduces to
\begin{align}
\alpha^{(1)}&=\norm{\bm\eta}_2^2I,\qquad
\beta^{(1)}=H_b+\bm\eta^\dagger\bm L+\bm L^\dagger\bm\eta.
\label{eq:SM_CE_first_order}
\end{align}

It remains to choose the jump coefficients $\bm\eta$ so that their ground--excited contribution in Eq.~\eqref{eq:SM_CE_first_order} cancels $H_b$. Each coefficient $\eta_j$ multiplies the decay vector $\ket{\ell_j}$, so this operator cancellation can be written as a vector equation in the excited manifold. Define the linear map
\begin{align}
\Lambda&:=\bigl(\ket{\ell_1},\ldots,\ket{\ell_J}\bigr),\qquad
\Lambda\bm\eta=\sum_j\eta_j\ket{\ell_j},\qquad
G=\Lambda\Lambda^\dagger.
\label{eq:SM_Lambda_definition}
\end{align}
Thus, $\Lambda$ maps the coefficient vector $\bm\eta\in\mathbb C^J$ to the corresponding superposition of physical decay vectors, and $\operatorname{im}\Lambda=\operatorname{im}G$. Since $\bm\eta^\dagger\bm L=\ket g\bm\eta^\dagger\Lambda^\dagger$ and $\bm L^\dagger\bm\eta=\Lambda\bm\eta\bra g$, Eq.~\eqref{eq:SM_CE_first_order} becomes
\begin{align}
\beta^{(1)}
&=\left(\Lambda\bm\eta+\frac i2\ket b\right)\bra g
+\ket g\left(\bm\eta^\dagger\Lambda^\dagger-\frac i2\bra b\right).
\label{eq:SM_beta_vector_form}
\end{align}
The two terms are adjoints, so $\beta^{(1)}=0$ when their excited-manifold coefficient vanishes. Because $\ket b\in\operatorname{im}G=\operatorname{im}\Lambda$, this cancellation equation has a solution. Since $\norm{\alpha^{(1)}}=\norm{\bm\eta}_2^2$, the tightest bound within this family uses the minimum-norm solution:
\begin{align}
\Lambda\bm\eta&=-\frac i2\ket b,\qquad
\bm\eta_\star=-\frac i2\Lambda^+\ket b.
\label{eq:SM_CE_cancellation}
\end{align}
The pseudoinverse identity $(\Lambda^+)^\dagger\Lambda^+=G^{-1}$ converts the minimum coefficient norm $\norm{\bm\eta_\star}_2$ into the inverse-relaxation geometry. This explicit choice of $\mathsf h(dt)$ therefore gives
\begin{align}
4\norm{\alpha^{(1)}}
&=4\norm{\bm\eta_\star}_2^2
=\bra bG^{-1}\ket b=R,\qquad
F_{\rm opt}(T)\le RT.
\label{eq:SM_universal_bound}
\end{align}
Thus, this choice yields the upper bound on the information rate identified in Sec.~\ref{sec:SM_multilevel}.

\section{Direct realization of the rate with fresh meters}
\label{sec:SM_direct}

This section derives the weak-measurement-and-feedback protocol that attains the CE rate. Building on the information-transfer mechanism of the amplitude-damping qubit~\cite{Kurdzialek2025}, we begin with an arbitrary monitored direction and show that stabilizing the multilevel rate-optimal response determines both this direction and the monitoring rate.

To derive one measurement--feedback step, consider a sensor state with reference state $\ket g$ at $\theta=0$ and response vector $\ket s$. Its local expansion is
\begin{align}
\rho_\theta
&=\proj g+\frac{\theta}{2}
\left(\ket s\bra g+\ket g\bra s\right)+O(\theta^2)
=\proj{\psi_\theta}+O(\theta^2),\qquad
\ket{\psi_\theta}:=\ket g+\frac{\theta}{2}\ket s.
\label{eq:SM_local_pure_representation}
\end{align}
The vector $\ket{\psi_\theta}$ is a local representative that reproduces the reference state and tangent through first order and simplifies the conditional Kraus calculation below.
Choose an arbitrary normalized direction $\ket v$ in the excited manifold and decompose the response as
\begin{align}
\ket s&=s_v\ket v+\ket{s_\perp},
\qquad
s_v:=\langle v|s\rangle,
\qquad
\langle v|s_\perp\rangle=0.
\label{eq:SM_arbitrary_monitor_direction}
\end{align}
The phase of $\ket v$ is chosen so that $s_v$ is real in the monitored quadrature.

Consider a weak measurement of the $g$--$v$ coherence. Define $X_v=\ket v\bra g+\ket g\bra v$, $Y_v=-i\ket g\bra v+i\ket v\bra g$, $P_{gv}=\proj g+\proj v$, and $P_\perp=I-P_{gv}$. A binary instrument and its outcome-dependent feedback are
\begin{align}
E_r&=\frac12(I+r\sin\phi X_v),\qquad
K_r=\frac1{\sqrt2}\left[P_\perp+\cos(\phi/2)P_{gv}
+r\sin(\phi/2)X_v\right],\qquad
V_r=e^{ir\phi Y_v/2},\qquad r=\pm1,
\label{eq:SM_multilevel_instrument}
\end{align}
where $K_r^\dagger K_r=E_r$. The measurement first displaces the reference state in an outcome-dependent direction,
\begin{align}
K_r\ket g
&=\frac1{\sqrt2}\left(\cos(\phi/2)\ket g
+r\sin(\phi/2)\ket v\right),
\qquad
V_rK_r\ket g=\frac1{\sqrt2}\ket g.
\label{eq:SM_reference_realignment}
\end{align}
Thus, $p_r(0)=1/2$, and the feedback $V_r$ returns both reference branches to $\ket g$ before the next sensing interval.

To determine what remains in the sensor, apply the same corrected operation to the tangent. The decomposition in Eq.~\eqref{eq:SM_arbitrary_monitor_direction} gives
\begin{align}
K_r\ket s
&=\frac1{\sqrt2}\left[
\ket{s_\perp}+s_v\cos(\phi/2)\ket v
+r s_v\sin(\phi/2)\ket g
\right],\quad
V_rK_r\ket s
=\frac1{\sqrt2}\left[
\ket{s_\perp}+s_v\cos\phi\ket v
+r s_v\sin\phi\ket g
\right].
\label{eq:SM_corrected_tangent}
\end{align}
Combining this result with Eq.~\eqref{eq:SM_local_pure_representation} gives
\begin{align}
p_r(\theta)
=\frac12+\frac{\theta}{2}r s_v\sin\phi+O(\theta^2),\qquad
\ket{\psi_{r,\theta}^{\rm corr}}
=\ket g+\frac{\theta}{2}
\left(\ket{s_\perp}+s_v\cos\phi\ket v\right)+O(\theta^2).
\label{eq:SM_corrected_branch}
\end{align}
The corrected sensor tangent is independent of $r$: feedback has aligned the branches, while the component along $\ket v$ has been reduced by $\cos\phi$. The same readout adds the classical FI
\begin{align}
\Delta F_{\rm rec}
&=\sum_{r=\pm1}\frac{[\partial_\theta p_r(0)]^2}{p_r(0)}
=s_v^2\sin^2\phi.
\label{eq:SM_single_readout_FI}
\end{align}
Consequently, one measurement--feedback step acts as
\begin{align}
s_v&\mapsto s_v\cos\phi,\qquad
\ket{s_\perp}\mapsto\ket{s_\perp},\qquad
\Delta F_{\rm rec}=s_v^2\sin^2\phi.
\label{eq:SM_multilevel_step}
\end{align}

Introduce a continuous monitoring rate $m\ge0$ through $\sin^2\phi=2m dt+o(dt)$. Since $\cos\phi=1-m dt+o(dt)$, the sensor response transforms as
\begin{align}
\ket s
\mapsto s_v\cos\phi\ket v+\ket{s_\perp}
&=\ket s-(1-\cos\phi)s_v\ket v
=\ket s-(1-\cos\phi)\proj v\ket s
=\ket s-m dt\proj v\ket s+o(dt) \\
&=(I-Mdt)\ket s+o(dt),
\qquad
M:=m\proj v.
\label{eq:SM_monitor_generator}
\end{align}
At the same time,
\begin{align}
\Delta F_{\rm rec}
&=2m|\langle v|s\rangle|^2dt+o(dt)
=2\bra sM\ket s dt+o(dt).
\label{eq:SM_record_increment}
\end{align}
Thus, the same positive rank-one operator $M$ describes the sensitivity removed from the sensor and the FI deposited in the record.

Free sensing and relaxation give $\frac d{dt}\ket s=\ket b-G\ket s/2$. To first order in $dt$, composing this evolution with the measurement--feedback step adds their response increments. The conditional FI of successive readouts accumulates in the record: each conditional score has zero mean, so cross terms between scores from distinct steps vanish~\cite{Gammelmark2014}. The two quantities therefore obey
\begin{align}
\frac d{dt}\ket s
&=\ket b-\frac12G\ket s-M\ket s,
\qquad
\dot F_{\rm rec}=2\bra sM\ket s.
\label{eq:SM_direct_pair_general}
\end{align}
The first equation tracks the sensitivity remaining in the sensor, and the second tracks the FI already transferred to the classical record. Adding the record-FI rate to the derivative of the sensor QFI cancels the monitoring terms. Using $\ket{s_\star}=G^{-1}\ket b$ and $G\ket{s_\star}=\ket b$ then gives
\begin{align}
\frac d{dt}\left(F_{\rm rec}+\norm{\ket s}^2\right)
&=2\operatorname{Re}\langle s|b\rangle-\bra sG\ket s
=R-(\bra s-\bra{s_\star})G(\ket s-\ket{s_\star})
\le R.
\label{eq:SM_information_balance}
\end{align}
Thus, monitoring redistributes accessible information between the sensor and the record, while the relaxation-weighted mismatch from $\ket{s_\star}$ is the deficit from the rate $R$. Asymptotic saturation follows whenever
\begin{align}
\lim_{T\to\infty}\frac1T\int_0^T
(\bra{s(t)}-\bra{s_\star})G(\ket{s(t)}-\ket{s_\star})dt&=0.
\label{eq:SM_saturation_criterion}
\end{align}
A direct way to realize this condition with a time-independent rank-one monitor is to make $\ket{s_\star}$ a fixed point of the controlled response dynamics. Requiring the right-hand side of Eq.~\eqref{eq:SM_direct_pair_general} to vanish at $\ket s=\ket{s_\star}$ gives
\begin{align}
0
&=\ket b-\frac12G\ket{s_\star}-M\ket{s_\star}
=\frac12\ket b-M\ket{s_\star},
\qquad
M\ket{s_\star}=\frac12\ket b.
\label{eq:SM_stationarity_condition}
\end{align}
For $M=m\proj v$, the vector $M\ket{s_\star}=m\ket v\langle v|s_\star\rangle$ lies along $\ket v$. Equation~\eqref{eq:SM_stationarity_condition} therefore forces the monitored direction to be parallel to $\ket b$. Since $\ket b$ is normalized, we may choose $\ket v=\ket b$ by fixing its phase. The component of the rate-optimal response along the bright direction is then
\begin{align}
\langle b|s_\star\rangle&=\bra bG^{-1}\ket b=R.
\label{eq:SM_bright_direction}
\end{align}
The remaining scalar condition determines the monitoring rate:
\begin{align}
m_\star&=\frac{1}{2R},
\qquad
M_\star=m_\star\proj b=\frac{\ket b\bra b}{2R}.
\label{eq:SM_Mstar_def}
\end{align}
Since $R=\tau_b$, the mean population lifetime of the bright state fixes the monitoring rate, $m_\star=1/(2\tau_b)$. Note that the single-step derivation assumed that the monitored component $\langle v|s\rangle$ is real. To verify that the fixed choice $\ket v=\ket b$ implements the same monitor throughout the controlled transient, let $A=G/2+M_\star$. Solving $\dot{\ket s}=\ket b-A\ket s$ from $\ket{s(0)}=0$ gives
\begin{align}
\ket{s(t)}
&=\int_0^t e^{-A(t-t')}\ket b\,dt'
=\int_0^t e^{-A\tau}\ket b\,d\tau.
\label{eq:SM_controlled_solution}
\end{align}
Because $A$ is Hermitian, $e^{-A\tau}$ is Hermitian for every $\tau$, and Eq.~\eqref{eq:SM_controlled_solution} therefore gives
\begin{align}
\langle b|s(t)\rangle
&=\int_0^t\langle b|e^{-A\tau}|b\rangle d\tau\in\mathbb R.
\label{eq:SM_fixed_quadrature}
\end{align}
Thus, a single fixed quadrature realizes the response equation for the entire protocol. The choice also satisfies $M_\star\ket{s_\star}=\ket b/2$, so signal generation is exactly balanced by relaxation and monitoring. Moreover,
\begin{align}
2\bra{s_\star}M_\star\ket{s_\star}&=R,
\label{eq:SM_steady_record_rate}
\end{align}
and the stationary sensor continually deposits FI into the record at the CE rate. The full rate-optimal response may span several decay modes, although the required monitor acts only on the single bright direction. Each meter is measured and discarded, and only its classical outcome persists.

\section{Convergence and constant-gap bound}
\label{sec:SM_transfer}

This section shows that the monitor $M_\star$ drives the response toward $\ket{s_\star}$ and that the transient mismatch in Eq.~\eqref{eq:SM_information_balance} has a finite integrated cost. Let $\ket\delta=\ket s-\ket{s_\star}$ and $A=G/2+M_\star$; since $\ket{s(0)}=0$, the initial mismatch is $\ket{\delta(0)}=-\ket{s_\star}$. The fixed-point condition in Eq.~\eqref{eq:SM_stationarity_condition} implies
\begin{align}
\frac d{dt}\ket\delta&=-A\ket\delta,
\qquad
\frac d{dt}\norm{\ket\delta}^2
=-\bra\delta(G+2M_\star)\ket\delta
\le-\bra\delta G\ket\delta,
\nonumber\\
\int_0^T\bra{\delta(t)}G\ket{\delta(t)}dt
&\le\norm{\ket{\delta(0)}}^2-\norm{\ket{\delta(T)}}^2
\le\norm{\ket{\delta(0)}}^2
=\norm{\ket{s_\star}}^2=:C.
\label{eq:SM_transient_cost}
\end{align}
This argument also covers singular $G$. Because $\ket{s_\star},\ket b\in\operatorname{im}G$ and both $G$ and $M_\star$ preserve $\operatorname{im}G$, the trajectory $\ket{\delta(t)}$ remains in $\operatorname{im}G$. On this finite-dimensional subspace, $G$ is positive definite, so $\ket{\delta(t)}$ converges exponentially to zero.

Integrating Eq.~\eqref{eq:SM_information_balance}, and using the classical--quantum decomposition in Eq.~\eqref{eq:SM_cq_QFI} for an optimal final sensor measurement, gives
\begin{align}
F_{\rm dir}(T)
&=F_{\rm rec}(T)+\norm{\ket{s(T)}}^2
=RT-\int_0^T\bra{\delta(t)}G\ket{\delta(t)}dt
\ge RT-C.
\label{eq:SM_direct_lower}
\end{align}
Together with the CE bound and the inclusion of the direct-monitoring protocol in the unrestricted class,
\begin{align}
RT-C&\le F_{\rm dir}(T)\le F_{\rm opt}(T)\le RT,\qquad
\lim_{T\to\infty}\frac{F_{\rm dir}(T)}T=\lim_{T\to\infty}\frac{F_{\rm opt}(T)}T=R.
\label{eq:SM_final_sandwich}
\end{align}
The direct-monitoring protocol therefore saturates the ultimate information rate up to a time-independent bound on the information deficit.

\section{Additivity of the ultimate rate and intersensor entanglement}
\label{sec:SM_many}

This section obtains the collective upper bound from the single-channel CE calculation and shows that independent local rank-one monitoring attains the resulting rate.

For $n$ identical sensors with independent relaxation, each interval $dt$ applies $\mathcal N_{\theta,dt}^{\otimes n}$, so time $T$ entails $N_{\rm use}=nT/dt$ uses of the same single-sensor channel. The local channels act on distinct subsystems and can be ordered sequentially within each interval. The adaptive $N$-use class in Eq.~\eqref{eq:SM_CE_discrete} therefore includes this parallel protocol, together with entangled inputs, collective controls, noiseless memories, feedback, and joint QEC~\cite{Demkowicz2017}.

Use the single-sensor Kraus choice constructed above, for which $\beta=O(dt^2)$ and $\alpha=\alpha^{(1)}dt+O(dt^2)$ with $4\norm{\alpha^{(1)}}=R$. Substituting $N=N_{\rm use}$ and $x=dt^{-1/2}$ in Eq.~\eqref{eq:SM_CE_discrete}, and taking $dt\to0$ at fixed $n$ and $T$, makes all $\beta$-dependent terms vanish and yields
\begin{align}
F_{\rm opt}^{(n)}(T)&\le 4nT\norm{\alpha^{(1)}}=nRT.
\label{eq:SM_copy_CE}
\end{align}

To attain this rate, initialize each sensor in $\ket g$ and apply the single-sensor rank-one protocol independently, using separate fresh meters, local feedback, and a final local sensor readout. The complete measurement records of different sensors are statistically independent at fixed $\theta$, so their FI adds. The single-sensor bound then gives
\begin{align}
F_{\rm dir}^{(n)}(T)&=nF_{\rm dir}(T)\ge nRT-nC.
\label{eq:SM_copy_lower}
\end{align}
Combining the achievable FI with the unrestricted upper bound gives
\begin{align}
nRT-nC&\le F_{\rm dir}^{(n)}(T)\le F_{\rm opt}^{(n)}(T)\le nRT,\qquad
\lim_{T\to\infty}\frac{F_{\rm dir}^{(n)}(T)}T
=\lim_{T\to\infty}\frac{F_{\rm opt}^{(n)}(T)}T=nR.
\label{eq:SM_copy_rate}
\end{align}
Thus, one local rank-one monitor acting along the bright direction of each sensor attains the unrestricted collective asymptotic rate, and intersensor entanglement cannot increase it.

\section{Numerical protocols and convergence checks}
\label{sec:SM_numerics}

This section specifies the comparison protocols in Fig.~2 of the main text. We use $G=\gamma_1\operatorname{diag}(1,9)$ in the decay-mode basis $\{\ket{\mu_1},\ket{\mu_2}\}$, $\ket b=(\ket{\mu_1}+\ket{\mu_2})/\sqrt2$, and units with $\gamma_1=1$. Every reported total FI includes the record and an optimal final sensor readout.

For free evolution with periodic readout and reset, each cycle starts in $\ket g$, evolves freely for time $\tau$, and ends with an optimal local readout followed by re-preparation of $\ket g$. Readout and reset take negligible time. For each stopping time $T$, we optimize $0<\tau\leq T$ to maximize the total FI, including the final incomplete cycle. We search the intervals with fixed $k=\lfloor T/\tau\rfloor$, including their endpoints. The bound $F_Q(t)\leq t^2$ implies $F_{\rm reset}(T;\tau)\leq T\tau$, allowing periods too short to improve the best value already found to be excluded. With the optimizing period denoted by $\tau=\tau_{\rm opt}(T)$, the plotted value is
\begin{align}
F_{\rm reset}(T)&=kF_Q(\tau)+F_Q(T-k\tau),\qquad k=\lfloor T/\tau\rfloor.
\label{eq:SM_reset_numerics}
\end{align}

For the direct-monitoring protocol, we integrate Eq.~\eqref{eq:SM_direct_pair_general} with $M=M_\star$. The open squares iterate the exact free-response step followed by the finite measurement--feedback map in Eq.~\eqref{eq:SM_multilevel_step}, using $\gamma_1dt=0.05$. This recursion is deterministic and therefore has no sampling error bars. Reducing the step to $\gamma_1dt=0.025$ and $0.0125$ confirms convergence to the continuous result.

For the no-feedback comparison, we use the same binary instrument along $\ket b$ without $V_r$. At each integer stopping time $1\leq\gamma_1T\leq40$, we optimize the total FI over a monitoring rate $m$ that remains constant during the run. We search the grid $m/\gamma_1=0,0.01,\ldots,0.60$, using 8000 trajectories at each nonzero rate. Each trajectory is propagated using the exact free channel and the normalized measurement update, together with their parameter derivatives. The total FI combines the accumulated conditional outcome FI with the average final sensor QFI, as in Eq.~\eqref{eq:SM_cq_QFI}. The plotted values are evaluated at the selected rates using an independent sample of $2\times10^4$ trajectories with $\gamma_1dt=0.02$. Error bars denote one Monte Carlo standard error. Reducing $\gamma_1dt$ to $0.01$ and $0.005$ at $\gamma_1T=5,10,20,40$ changes $F(T)/(RT)$ by at most $0.0034$.

\bibliography{reference}